\documentstyle[11pt]{article}
\def\v1{\vspace{1cm}}
\def\be{\begin{equation}}
\def\ee{\end{equation}}
\def\bc{\begin{center}}
\def\ec{\end{center}}

\newcommand{\bea}{\begin{eqnarray}}
\newcommand{\eea}{\end{eqnarray}}

\begin{document}

\title
{\bf Time-reparametrization-invariant dynamics of a relativistic string}
\author{
B.M. Barbashov, V.N. Pervushin \\
{\normalsize\it Joint Institute for Nuclear Research},\\
 {\normalsize\it 141980, Dubna, Russia.}\\
}

\date{\empty}

\maketitle

\medskip
PACS number(s):11.25.-w
\medskip


\begin{abstract}
{\small
{
The time-reparametrization-invariant dynamics of a relativistic string
is studied in the Dirac generalized Hamiltonian theory
by resolving the first class constraints.
The reparametrization-invariant evolution parameter is identified with
the time-like coordinate of the "center of mass"  of a string which
is separated from local degrees of freedom by transformations
conserving the group of diffeomorphisms of
the generalized Hamiltonian formulation and the Poincare covariance
of local constraints.
To identify the "center of mass" time-like coordinate with the invariant
proper time (measured by an observer in the comoving frame of reference),
 we apply
the Levi-Civita - Shanmugadhasan canonical transformations which convert
the global (mass-shell) constraint into a new momentum, so that the
corresponding gauge is not needed for the Hamiltonian reduction.

The resolving of local
constraints leads to an "equivalent unconstrained system" of the type of
the R\"ohrlich string. Our classical Hamiltonian formalism naturally
provides this approach to quantum theory of relativistic string.
}}

\end{abstract}



\section{Introduction}

The group of diffeomorphisms of the Hamiltonian description
of relativistic systems (particles, string, branes, general relativity)
~\cite{ADM}-\cite{adm} contains the Abelian subgroup of
the reparametrization of the coordinate time~\cite{vlad}.
All known descriptions of a relativistic string~\cite{gsw,bn}
are based on the reduction of the extended phase space by the
fixation of gauge~\cite{ty,hrt}
which breaks reparametrization - invariance from very beginning.
The questions arise: Can one describe
the time - reparametrization - invariant dynamics of a relativistic string
dynamics directly in the terms of
reparametrization - invariant variables, and what is a difference of
this description from the gauge-fixing method?

To answer these questions,
in the present paper, we apply a method
of a reparametrization - invariant Hamiltonian description
developed for gravitation~\cite{kuchar,grg,ps1,pp}.

The method of a reparametrization - invariant description is based on
the reduction of an action by the explicit resolving of the first
class constraints.
An important element of the invariant reduction is
the Levi-Civita - Shanmugadhasan canonical transformation~\cite{lc,sh}
that linearizes the energy constraint as the generator
of reparametrizations of the coordinate time.

The content of the paper is the following.
In Section 2 we formulate  the method of
the invariant Hamiltonian reduction using the simplest examples of
classical mechanics and relativistic particle. Section 3 is devoted to
the generalized Hamiltonian approach to
 a relativistic string and the statement of the problem.
In Section 4, local excitations are separated from the "center of mass"
coordinates of the string.
In Section 5, the Levi-Civita transformations and
the invariant Hamiltonian reduction are performed to resolve the
global constraint and
to convert the time-like variable of the global motion
into the proper time.
In Section 6, the classical and quantum
dynamics of local excitations are described in terms of the proper time.
Section 7 is devoted to the generating functional for the Green functions.

\section{Invariant Hamiltonian Reduction}

\subsection{ Mechanics}

To illustrate the time-reparametrization-invariant Hamiltonian
reduction~\cite{grg} and its difference from
the gauge-fixing method, let us consider an extended form of
a classical-mechanical system
\be \label{mec}
W=\int\limits_{\tau^1 }^{\tau_2 }d\tau \left( p\dot q - \Pi_0 \dot Q_0
- \lambda [-\Pi_0 + H(p,q)]\right)~,
\ee
that is invariant under reparametrizations of the coordinate
evolution parameter $\tau$ and "lapse" function $\lambda$
\be \label{coo}
\tau \rightarrow \tau'=\tau'(\tau),
~~~~~\lambda~\rightarrow~\lambda'=\lambda \frac{d\tau}{d\tau'}~.
\ee
The problem of the classical description
is to obtain the evolution of the physical variables of
the {\it world space} $q, Q_0$
in terms of the {\it geometric time} $T$ defined as
\be \label{prom}
dT:=\lambda d\tau,~~~
~T=\int\limits_{0 }^{\tau } d\tau'\lambda(\tau')~,
\ee
that is also invariant under reparametrizations~(\ref{coo}).

The second problem (connected with quantization)
is to present the effective action
of the equivalent unconstrained theory
directly in terms of $T$, the equations of which reproduce this evolution.
The solution of the second problem
will be called the {\it invariant Hamiltonian reduction}.

The resolving of the first problem for the considered system
is trivial, as the equations of motion of this system
\be \label{eeq}
\dot q = \lambda \partial_p H,~~~~~
\dot p = - \lambda \partial_q H,~~~~ \dot Q_0 = \lambda,~~~~~\dot \Pi_0=0
\ee
in terms of the geomeric time~(\ref{prom})
\be \label{geq}
\frac{d q}{d T} =  \partial_p H,~~~~~
\frac{d p}{d T} = - \partial_q H,~~~~~ \frac{d Q_0}{d T} = 1,~~~~
\frac{d \Pi_0}{d T} =0
\ee
are completely equivalent to the equations of
the conventional unconstrained mechanics in the {\it reduced
phase space} $(p,q)$
\be \label{rmec}
W_{reduced}=\int\limits_{T(\tau_1)=T_1 }^{T(\tau_2)=T_2 }dT \left( p\frac{d q}{dT}
 -  H(p,q)\right).
\ee
The problem is how to derive this system from the extended one~(\ref{mec})
to apply the simplest Hamiltonian quantization with a clear physical
interpretation of the invariant quantities.

The solution of the problem of the {\it invariant Hamiltonian reduction}
considered in the present review
is the explicit resolving of three equations of the extended
system~(\ref{mec}):\\
i) for the variable $\lambda$ (treated as constraint)
\be \label{mc}
\frac{\delta W}{\delta\lambda}=-\Pi_0+H(p,q)=0~,
\ee
ii) for the  momentum $\Pi_0$ with a negative contribution to
the constraint~(\ref{mc})
\be \label{moment}
\frac{\delta W}{\delta\Pi_0}=0~\Rightarrow~ \frac{dQ_0}{d\tau}=\lambda~,
\ee
and iii) for its conjugate variable $Q_0$
\be \label{dep}
\frac{\delta W}{\delta Q_0}= \frac{d\Pi_0}{d\tau}=0~.
\ee
(We call these three equations ~(\ref{mc}) -~(\ref{dep})
the {\it geometric sector}.)

The resolving of the constraint~(\ref{mc}) expresses
the "ignorable" momentum $\Pi_0$
through $H(p,q)$ with a positive value $\Pi_0=H(p,q) > 0$.
The second equation~(\ref{moment}) identifies the
{\it dynamic evolution parameter} $Q_0$ with the proper time~(\ref{prom})
$Q_0 = T$.
It is not the gauge but the invariant solution of the equation of
motion~(\ref{moment}). The third equation~(\ref{dep}) is the conservation law.

As a result of the invariant Hamiltonian
reduction (i.e., a result of the substitution
of $\Pi_0=H$ and $Q_0=T$ into the initial action~(\ref{mec})~) this action
is reduced to the one of the conventional mechanics~(\ref{rmec})
in terms of the proper time $T$ where
the role of the nonzero Hamiltonian
of evolution in the proper time $T$ is played by
the constraint-shell value of the "ignorable" momentum $\Pi_0=H(p,q)$.
In other words, this constraint-shell action
$W(\mbox{constraint})=W^M$
determines the nonzero  Hamiltonian $H(p,q)$ in the proper time $T$,
instead of the zero generalized Hamiltonian in
the coordinate time $\tau$ in~(\ref{mec}) $\lambda(-\Pi_0+H)$.

Thus, the equivalent unconstrained system
was constructed without any additional constraint of the type:
\be \label{gem}
\lambda = 1 ,~~~~~~~~~~~\tau = T
\ee
which confuse quantities of the measurable sector
with noninvariant ones. This confusion is contradictable.
The "gauge-fixing" identification of the coordinate evolution parameter
$\tau$ and the geometric time $dT=\lambda d\tau$ in the form of the gauges
(\ref{gem}) contradicts to
the difference of their Hamiltonians $\lambda (-\Pi_0+H)\not= H(p,q)$.

The second difference of the "gauge-fixing" from
the invariant Hamiltonian reduction is more essential, namely,
the formulation of the theory in terms of the invariant geometric
time~(\ref{prom}) is achieved by the explicit resolving of the
constraint~(\ref{mc}) and equation of motion~(\ref{moment}),
as a result of which "ignorable" variables $\Pi_0,Q_0$ are excluded
from the phase space.

\subsection{Special Relativity}

Let us apply the invariant Hamiltonian reduction
to relativistic particle.

To answer the question: Why is the reparametrization-invariant
reduction needed?, let us consider
 relativistic mechanics in the Hamiltonian form~\cite{grg}
\be \label{SR}
W[P,X|N|\tau_1,\tau_2]=
\int\limits_{\tau_1 }^{\tau_2 }d\tau [- P_{\mu}\dot X^{\mu}
- \frac{N}{2m}(-P_{\mu}^2+m^2) ]~,
\ee
which is classically equivalent to the conventional square root form
\be \label{srsr}
W[X|\tau_1,\tau_2]=
-m\int\limits_{\tau_1 }^{\tau_2 }d\tau \sqrt{\dot X^{\mu}\dot X_{\mu}}
\ee
Both these action is invariant with respect to reparametrizations of
the {\it coordinate evolution parameter} \be \label{coo1}
\tau \rightarrow \tau'=\tau'(\tau),~~~~~N'd\tau'=Nd\tau
\ee
given in the one-dimensional space with the invariant interval
\be \label{pros}
dT:=Nd\tau,~~~~~~~~T=\int\limits_{0 }^{\tau } d\bar \tau N(\bar \tau)
\ee
We called this invariant interval
the {\it geometric time}~\cite{grg} whereas
the dynamic variable $X_0$ (with a negative contribution
in the constraint) we called {\it dynamic evolution parameter}.

In terms of the geometric time~(\ref{pros})
the classical equations of the generalized Hamiltonian system~(\ref{SR})
takes the form
\be \label{cesr}
\frac{d X_{\mu}}{d T}=\frac{P_{\mu}}{m},~~~~~~~
\frac{d P_{\mu}}{d T}=0,~~~~~~~P_{\mu}^2-m^2=0.
\ee
The classical problem is to find the evolution of the world space
variables with respect to the geometric time $T$.

The quantum problem is
to obtain the equivalent unconstrained theories
directly in terms of the invariant times $X_0$ or $T$ with
the invariant Hamiltonians of evolution.
The solution of the second problem is called
the dynamic (for $X_0$), or geometric
(for $T$) {\it reparametrization-invariant Hamiltonian reductions}.

The dynamic reduction of the action~(\ref{SR}) means the substitution
of the explicit resolving of the energy constraint $(-P_{\mu}^2+m^2)=0$ with
respect to the momentum $P_0$ into this action
\be \label{po}
\frac{\delta W}{\delta N}=0~~\Rightarrow~ ~P_0=\pm\sqrt{m^2+P_i^2}.
\ee
In accordance with two signs of the solution~(\ref{po}),
after the substitution of~(\ref{po}) into~(\ref{SR}), we have
two branches of the dynamic unconstrained system
\be \label{srd}
W(\mbox{constraint})_{\pm}=
\int\limits_{X_0(\tau_1)=X_0(1)}^{X_0(\tau_2)=X_0(2)} dX_0
\left[ P_{i}\frac{dX_i}{dX_0}\mp\sqrt{m^2+P_i^2} \right]~.
\ee
The role of the time of evolution, in this action,
is played by the variable $X_0$ that abandons the Dirac sector of
"observables" $P_i, X_i$, but not the sector of
"measurable" quantities.
At the same time, its conjugate momentum $P_0$
converts into the corresponding Hamiltonian of evolution,
values of which are energies of a particle.

This invariant reduction of the action gives an "equivalent"  unconstrained
system together with definition of the dynamic evolution parameter
$X_0$ corresponding to a nonzero Hamiltonian $P_0$.

Thus, we need the reparametrization-invariant Hamiltonian reduction to
determine the dynamic evolution parameter and its invariant Hamiltonian for
a reparametrization-invariant system and to apply the symplest
canonical quantization to it.

In quantum relativistic theory, we get two Schr\"odinger equations
\be \label{srw}
i\frac{d}{dX_0}\Psi_{(\pm)}(X|P)=\pm\sqrt{m^2+P_i^2}\Psi_{(\pm)}(X|P)~,
\ee
with positive and negative values of $P_0$
and normalized wave functions
\be \label{srnw}
\Psi_{\pm}(X|P)
=\frac{A_P^{\pm}\theta(\pm P_0)}{(2\pi)^{3/2}\sqrt{2P_0}}
\exp(-iP_{\mu}X^{\mu}),
~~~~~~~
\left([A_P^{-},A_{P'}^{+}]=\delta^3(P_i-P'_i)\right)~.
\ee
The coefficient $A_P^{+}$, in the secondary
quantization, is treated as the operator of creation
of a particle with positive energy;
and the coefficient $A_P^{-}$, as the operator of annihilation  of
a particle also with positive energy.
The physical states are formed by action of these operators on the vacuum
$<0|,|0>$ in the form of out-state (~$|P>=A_P^+|0>$~) with
positive frequencies
and in-state (~$<P|=<0|A_P^-$~) with
negative frequencies.
This treatment means that positive frequencies propagate forward
(${X_0}_2>{X_0}_1$);
and negative frequencies, backward (${X_0}_1>{X_0}_2$), so that
the negative values of energy are excluded from the spectrum
to provide the stability of the quantum system
in QFT~\cite{bww}. For this causal convention the geometric time~(\ref{pros})
is always positive in accordance with the equations of motion~(\ref{cesr})
\be \label{at}
\left(\frac{d T}{d X_0}\right)_{\pm}=\pm \frac{m}{\sqrt{P_i^2+m^2}}~~
\Rightarrow~~T({X_0}_2,{X_0}_1)=
\pm \frac{m}{\sqrt{P_i^2+m^2}}({X_0}_2-{X_0}_1) \geq 0
\ee
In other words, instead of  changing the sign of energy,
we change that of the dynamic evolution parameter, which leads to
the arrow of the geometric time~(\ref{at}) and to the causal Green function
\be \label{caus}
G^c(X)=G_+(X)\theta(X_0)+G_-(X)\theta(-X_0)=
i\int\limits_{ }^{ }\frac{d^4P}{(2\pi)^4}
\exp(-iPX)\frac{1}{P^2-m^2-i\epsilon},
\ee
where $G_+(X)=G_-(-X)$ is the "commutative" Green function~\cite{bww}
\be \label{FIp}
G_{+}(X)=\int\limits_{ }^{ }\frac{d^4P}{(2\pi)^3}
\exp(-iPX)\delta(P^2-m^2)\theta(P_0)=
\ee
$$
\frac{1}{2\pi}\int\limits_{ }^{ }d^3Pd^3P'
<0|\Psi_{-}(X|P)\Psi_{+}(0|P')|0>~.
$$

The question appears: How to construct the path integral without
gauges?

To obtain the reparametrization-invariant form of the functional integral
adequate to the considered gauge-less reduction~(\ref{srd}) and the
causal Green function~(\ref{caus}), we use
the version of composition law for the commutative Green function
with the integration over the whole measurable sector $X_{1 \mu}$
\be \label{Dl}
G_+(X-X_0)=\int\limits_{ }^{ } G_+(X-X_1)\bar G_+(X_1-X_0)dX_1~,~~~~~~~~
~\bar G_+=\frac{G_+}{2\pi\delta(0)}~,
\ee
where $\delta(0)=\int dN$ is the infinite volume of the group
of reparametrizations of the coordinate $\tau$.
Using the composition law $n$-times, we got
the multiple integral
\be \label{firp}
G_+(X-X_0)=\int\limits_{ }^{ }G_+(X-X_1)
\prod\limits_{k=1}^{n}\bar G_+(X_k-X_{k+1})dX_k~,~~~~~~~
(~X_{n+1}=X_0~)~.
\ee
The continual limit of the multiple integral with the integral representation
for $\delta$-function
$$
\delta(P^2-m^2)=\frac{1}{2\pi}\int\limits_{ }^{ } d N \exp[i N (P^2-m^2)]
$$
can be defined as the path integral
in the form of the average over the group of reparametrizations
\be \label{srfi}
G_+(X)=\int\limits_{X(\tau_1)=0 }^{X(\tau_2)=X }
\frac{dN(\tau_2)d^4P(\tau_2)}{(2\pi)^3}
\prod\limits_{\tau_1 \leq \tau < \tau_2}^{ }\left\{ d\bar N(\tau)
\prod\limits_{\mu}\left( \frac{dP_{\mu}(\tau)dX_{\mu}(\tau)}{2\pi}\right)
\right\}
\ee
$$
\exp(i W[P,X|N|\tau_1,\tau_2]),
$$
where $\bar N={N}/{2\pi\delta(0)}$,
and $W$ is the initial extended action~(\ref{SR}).

\subsection{Geometric unconstrained system for a relativistic particle}

The Hamiltonian of the unconstrained system in terms of the geometric time $T$
can be obtained by the canonical Levi-Civita - type
transformation~\cite{lc,sh,gkp1}
\be
(P_{\mu}, X_{\mu}) \Rightarrow\, (\Pi_{\mu}, Q_{\mu})
\ee
to the variables ($\Pi_{\mu},Q_{\mu}$) for which one of equations
identifies $Q_0$ with the geometric time $T$.
This transformation
converts the constraint into a new momentum
\be \label{levi}
\Pi_0= \frac{1}{2m}[P_{0}^2 - P_i^2] ,~~~~~~\Pi_i=P_i,~~~~
 Q_0=X_0\frac{m}{P_0} ,~~~~~Q_i=X_i-X_0\frac{P_i}{P_0}
\ee
and has the inverted form
\be \label{ivel}
P_0=\pm \sqrt{2m\Pi_{0}+\Pi_i^2},~P_i=\Pi_i,~
X_0=\pm Q_0\frac{\sqrt{2m\Pi_{0}+\Pi_i^2}}{m},
~~X_i=Q_i+Q_0\frac{\Pi_i}{m}.
\ee
After transformation~(\ref{levi}) the action~(\ref{SR}) takes the form
\be \label{SRlc}
 W=\int\limits_{\tau_1}^{\tau_2} d\tau
\left[
- \Pi_{\mu}\dot Q^{\mu}- N(-\Pi_0+  \frac{m}{2} )-\frac{d}{d\tau}S^{lc}
\right],~~~~S^{lc}=(Q_0 \Pi_0).
\ee
The invariant reduction is the resolving of the constraint $\Pi_0={m}/{2}$
which determines a new Hamiltonian
of evolution with respect to the new dynamic evolution parameter $Q_0$,
whereas the equation of motion for this momentum $\Pi_0$
identifies  the dynamic evolution parameter $Q_0$ with the geometric time $T$
($dQ_0=dT$).
The substitution of these solutions into the action~(\ref{SRlc})
leads to the reduced action of a geometric unconstrained system
\be \label{geo}
W(\mbox{constraint})=\int\limits_{T_1}^{T_2} dT
\left( \Pi_{i}\frac{dQ_i}{dT}- \frac{m}{2} - \frac{d}{dT} (S^{lc})  \right)
~~~~~(S^{lc}=Q_0\frac{m}{2}),
\ee
where variables $\Pi_i,Q_i$ are cyclic ones and have the meaning
of initial conditions in the comoving frame
\be
 \frac{\delta W}{\delta \Pi_i}=
\frac{dQ_i}{d\tau}=0
\Rightarrow\,
Q_i=Q_i^{(0)},~~~~
\frac{\delta W}{\delta Q_i}=
\frac{d\Pi_i}{d\tau}=0
\Rightarrow\,
\Pi_i=\Pi_i^{(0)}.
\ee
The substitution of all geometric solutions
\be
Q_0=T,~~\Pi_0=\frac{m}{2},~~\Pi_i=\Pi_i^{(0)}=P_i,~~Q_i=Q_i^{(0)}
\ee
into the inverted Levi-Civita transformation~(\ref{ivel}) leads to the
conventional  relativistic solution for the dynamical system
\be \label{line}
P_0=\pm \sqrt{m^2+P_i^2},
~~~~P_i=\Pi_i^{(0)},~~~~
X_0(T)= T\frac{P_0}{m},~~~~
X_i(T)=X_i^{(0)}+T\frac{P_i}{m}.
\ee
The Schr\"odinger equation for the wave function
\be \label{geom}
\frac{d}{idT}\Psi (T,Q_i|\Pi_i)= \frac{m}{2}\Psi (T,Q_i|\Pi_i),~
\ee
$$
\Psi (T,Q_i|\Pi_i)=  \exp(-iT\frac{m}{2})
\exp(i\Pi_i^{(0)}Q_i)
$$
contains only one eigenvalue $m/2$  degenerated with
respect to the cyclic momentum $\Pi_i$.
We see that there are differences between the dynamic and geometric
descriptions.
The dynamic evolution parameter is given in the whole region $-\infty < X_0 < +\infty$,
whereas the geometric one is only positive $0<T< +\infty$, as it
follows from the properties of the causal Green function~(\ref{caus})
after the Levi-Civita transformation~(\ref{levi})
$$
G^c(Q_{\mu})=\int\limits_{-\infty }^{+\infty }d^4\Pi_{\mu}
\frac{\exp(iQ^{\mu}\Pi_{\mu})}{2m(\Pi_0-m/2-i\epsilon/2m)}=
\frac{\delta^3(Q)}{2m}\theta(T),~~~~~~T=Q_0~.
$$
Two solutions of the constraint (a particle and antiparticle) in the dynamic
system correspond to a single solution in the geometric system.

Thus, the reparametrization-invariant content of the equations of motion of a
relativistic particle in terms of the geometric time is covered
by two "equivalent" unconstrained systems: the dynamic and geometric. In both
the systems, the invariant times are not {\sl the coordinate evolution
parameter}, but variables with the negative contribution into the energy
constraint. The Hamiltonian description of a relativistic particle in terms
of the geometric time can be achieved by the Levi-Civita-type canonical
transformation, so that the energy constraint converts into a new momentum.
Whereas, the dynamic unconstrained system is suit for the secondary
quantization and the derivation of the causal Green function that determine
the arrow of the geometric time.

\section{Relativistic String}

\subsection{The generalized Hamiltonian formulation}

We begin with the action for a relativistic string
in the geometrical form~\cite{bvh}
\be \label{E}
W=-\frac{\gamma}{2}\int\limits_{}^{} d^2u \sqrt{-g} g^{\alpha\beta}
\partial_{\alpha}x^{\mu}\partial_{\beta}x_{\mu},~~~~u_{\alpha}=(u_0,u_1)
\ee
where the variables $x_{\mu}$ are string coordinates
given in a space-time with a dimension $D$
and the metric $(x_{\mu}x^{\mu}:=x_0^2-x_i^2)$;
$g_{\alpha\beta}$   is a second-rank metric tensor
given in
the two-dimensional Riemannian space $u_{\alpha}=(u_0,u_1)$.

To formulate the Hamiltonian approach, one needs to separate
the two-dimensional Riemannian space $u_{\alpha}=(u_0,u_1)$
on the set of space-like lines $\tau=\rm{constant}$ in the form of
the Dirac-Arnovitt-Deser-Misner parametrization of the
two-dimensional metric
\be \label{metr}
g_{\alpha,\beta}=\Omega^2
\left( \begin{array}{cc}\lambda_1^2-\lambda_2^2&\lambda_2\\\lambda_2&-1
\end{array}\right),~~~~~\sqrt{-g}=\Omega^2\lambda_1
\ee
with the invariant
interval~\cite{ADM}
\be \label{m}
ds^2 = g_{\alpha\beta} du^{\alpha} du^{\beta} =
\Omega^2[\lambda_1^2d\tau^2 - (d\sigma+\lambda_2d\tau)^2]~,
~~~~~u_{\alpha}=(u_0=\tau,u_1=\sigma)
\ee
where $\lambda_1$ and $\lambda_2 $ are known in general relativity (GR)
as the lapse function and shift " vector",
respectively~\cite{Y,bh}.
The action~(\ref{E})  after the substitution~(\ref{m})
does not depend on the conformal factor
$\Omega$ and takes the form
\be \label{P}
W=-\frac{\gamma}{2}\int\limits_{\tau_1}^{\tau_2} d\tau
\int\limits_{\sigma_1(\tau) }^{\sigma_2(\tau) }
d\sigma\left[\frac{(D_{\tau}x)^2}{\lambda_1} -
\lambda_1 x'^2 \right]
\ee
where
\be \label{cd}
D_{\tau}x_{\mu}=\dot x_{\mu}- \lambda_2 x'_{\mu}~~~~~~~~
(\dot x=\partial_{\tau}x,~ x'=\partial_{\sigma}x )
\ee
is the covariant derivative with respect to
the two-dimensional metric~(\ref{m}). The  metric~(\ref{m}),
the action~(\ref{P}), and the covariant derivative~(\ref{cd})
are invariant under the transformations (see Appendix A)
\be \label{kt}
\tau \Rightarrow\,\widetilde{\tau}=f_1(\tau),~~~~~~~~~~~~
\sigma \Rightarrow\,\widetilde{\sigma}=f_2(\tau,\sigma).
\ee
A similar group of transformations in GR is well-known
as the "kinemetric" group
of diffeomorphisms of the Hamiltonian
description~\cite{vlad}.

The variation of action~(\ref{P}) with respect to $\lambda_1$ and $\lambda_2$
leads to the equations
\be \label{lambda2}
\frac{\delta W}{\delta\lambda_2}=\frac{x'D_{\tau}x}{\lambda_1}=0~
\Rightarrow~\lambda_2=\frac{\dot x x'}{x'^2};~~~~~~~
\ee
$$
\frac{\delta W}{\delta\lambda_1}=\frac{(D_{\tau}x)^2}{\lambda_1^2}+x'^2=0~
\Rightarrow~~\lambda_1^2=\frac{(\dot x x')^2-\dot x^2 x'^2}{(x'^2)^2}
$$
The solutions of these equations
convert the action~(\ref{P}) into the standard Nambu-Gotto action of
a relativistic string~\cite{bn,rc}
$$
W=-\gamma\int\limits_{\tau_1}^{\tau_2} d\tau
\int\limits_{\sigma_1(\tau) }^{\sigma_2(\tau) }d\sigma
\sqrt{(\dot x x')^2 - \dot x^2 x'^2}.
$$
The generalized Hamiltonian
form~\cite{d2} is obtained by the Legendre transformation~\cite{gt} of
the action~(\ref{P})
\be   \label{kwe}
W=
\int\limits_{\tau_1}^{\tau_2} d\tau
\int\limits_{\sigma_1(\tau) }^{\sigma_2(\tau) }
d\sigma \left( -p_{\mu}D_{\tau} x^{\mu}
+\lambda_1 \phi_1  \right) ~=~
\int\limits_{\tau_1}^{\tau_2} d\tau
\int\limits_{\sigma_1(\tau) }^{\sigma_2(\tau) }d\sigma
\left( -p_{\mu}\dot x^{\mu} +\lambda_1 \phi_1
+ \lambda_2 \phi_2 \right),
\ee
where
\be   \label{h}
\phi_1= \frac{1}{2\gamma}[p_{\mu}^2+(\gamma x'_{\mu})^2],~~~~~~~~~~~~~
\phi_2 = x'^{\mu} p_{\mu},
\ee
and the generalized Hamiltonian
\be \label{ham}
{\cal H}=\lambda_1 \phi_1 + \lambda_2 \phi_2
\ee
is treated as the generator of evolution with respect to the coordinate time
$\tau$, and $ \lambda_1, \lambda_2$ play the role of variables with
the zero momenta
\be \label{h1}
P_{\lambda_1}=0,~~~~~~~~P_{\lambda_2}=0
\ee
considered as the first class primary constraints~\cite{d2,gt}.
The equations for $ \lambda_1, \lambda_2$
\be \label{cp}
\frac{\delta W}{\delta \lambda_1}=
\phi_1 = 0;~~~~~~~~~~~~~~~~~~~
\frac{\delta W}{\delta \lambda_2} =
\phi_2 = 0
\ee
are known as the first class secondary constraints~\cite{d2,hrt,gt}.
The Hamiltonian equations of motion take the form
\be \label{ex}
\frac{\delta W}{\delta x^{\mu}}=
\dot p_{\mu}-\partial_{\sigma}[\gamma\lambda_1 x'_{\mu}+\lambda_2 p_{\mu}]=0,
~~~~~~~~~~\frac{\delta W}{\delta p^{\mu}}=
p_{\mu}+
\gamma\frac{D_{\tau}x_{\mu}}{\lambda_1}=0
\ee

The problem is to find solutions of the Hamiltonian
equations of motion~(\ref{ex})
and constraints~(\ref{cp}) which are invariant with respect to the kinemetric
transformations~(\ref{kt}).

There is the problem of the solution of the linearized
"gauge-fixing" equation in terms of the evolution
parameter $\tau$ (as the object reparametrizations in the initial theory)
being adequate to the initial kinemetric invariant and relativistic invariant
system.
In particular, the constraints mix
 the global motion of the "center of mass" coordinates with
local excitations of a string $\xi_{\mu}$, which
contradicts to the relativistic invariance of internal degrees of freedom of
a string.
In this context, it is worth to clear up a set of questions:
Is it possible to introduce the reparametrization-invariant evolution
parameter for the string dynamics, instead of
the non-invariant coordinate time $(\tau)$ used as the evolution parameter
in the gauge-fixing method?
Is it possible to construct the observable nonzero Hamiltonian of
evolution of the "center of mass" coordinates?
What is relation of the "center of mass"  evolution to
the unitary representations of the Poincare group?

\section{ The separation of the "center of mass" coordinates}

To apply the reparametrization-invariant Hamiltonian reduction discussed before
to a relativistic string, one should define the {\it proper time}
in the form of the reparametrization-invariant functional of the
lapse function (of type~(\ref{pros})),
and to point out, among the variables, a {\it dynamic evolution parameter},
the equation of which identifies it with the proper time of
type~(\ref{moment}). As any extended object admits to define
the coordinates of its center of mass,
we identify this {\it dynamic evolution parameter} with the time-like
coordinate of the center of mass of a string
\be \label{X1}
  X_{\mu}(\tau)=
\frac{1}{l(\tau)}\int\limits_{\sigma_1(\tau) }^{\sigma_2(\tau) }
d\sigma x_{\mu}(\tau,\sigma),~~~~~~
l(\tau)=\sigma_2(\tau)-\sigma_1(\tau).
\ee
We see that the invariant reduction requires to separate the "center of mass"
variables before variation of the action.
This separation is fulfilled by the substitution of
\be \label{zh}
x_{\mu}(\tau,~\sigma)=
X_{\mu}(\tau)+\xi_{\mu}(\tau,~\sigma)
\ee
into the action~(\ref{P}), which takes the form
\be \label{cov}
W=-\frac{\gamma}{2}\int\limits_{\tau_1 }^{\tau_2 }d\tau \left\{
 \frac{\dot X^2 l(\tau)}{N_0(\tau)}+
2\dot X_{\mu}
\int\limits_{\sigma_1(\tau) }^{\sigma_2(\tau) }d\sigma
\frac{D_{\tau}\xi^{\mu}}{\lambda_1} +
\int\limits_{\sigma_1(\tau) }^{\sigma_2(\tau) }d\sigma
\left(\frac{(D_{\tau}\xi)^2}{\lambda_1}-\lambda_1\xi'^2\right)\right\},
\ee
where the global lapse function $N_0(\tau)$ is defined as the functional
of $\lambda_1(\tau,\sigma)$
\be \label{L}
\frac{1}{N_0[\lambda_1]}=
\frac{1}{l(\tau)}\int\limits_{\sigma_1(\tau) }^{\sigma_2(\tau) }
d\sigma \frac{1}{\lambda_1(\tau,\sigma)}.
\ee
>From definition~(\ref{X1}) and equality~(\ref{zh}) it follows that
the local variables $\xi_{\mu}$ are given
in the class of functions (with the nonzero Fourier harmonics) which
satisfy the conditions
\be \label{lhp}
\int\limits_{\sigma_1(\tau) }^{\sigma_2(\tau) }d\sigma \xi_{\mu}
(\tau,\sigma)=0.
\ee
The formulation of the Hamiltonian approach (consistent with~(\ref{X1}))
supposes the similar separation of the conjugate momenta $p_{\mu}$
defined by equation ~(\ref{ex}). If we substitute the definition~(\ref{zh})
in these equations, we get
\be \label{àzh}
p_{\mu}(\tau,~\sigma)=
-\gamma \left(\frac{{\dot X}_{\mu}(\tau)}{\lambda_1}+
\frac{D_{\tau}\xi_{\mu}(\tau,~\sigma)}{\lambda_1}\right).
\ee
Defining the total momentum of a string $P_{\mu}$
\be \label{lhp2}
P_{\mu}=
\int\limits_{\sigma_1(\tau) }^{\sigma_2(\tau) }
d\sigma p_{\mu}(\tau,\sigma)=-\gamma
\int\limits_{\sigma_1(\tau) }^{\sigma_2(\tau) }d\sigma
\left(\frac{{\dot X}_{\mu}(\tau)}{\lambda_1}+
\frac{D_{\tau}\xi_{\mu}(\tau,~\sigma)}{\lambda_1}\right),
\ee
and taking into account (\ref{L}) we obtain  the following expresion
\be \label{lhp3}
P_{\mu}=
-\gamma \frac{\dot X_{\mu}l}{N_0(\tau)}-
\gamma \int\limits_{\sigma_1(\tau) }^{\sigma_2(\tau) }d\sigma
\frac{D_{\tau}\xi_{\mu}(\tau,~\sigma)}{\lambda_1},
\ee
therefore the equality
\be \label{P1}
\int\limits_{\sigma_1(\tau) }^{\sigma_2(\tau) }d\sigma \frac{D_{\tau}
\xi^{\mu}}{\lambda_1}=
\int\limits_{\sigma_1(\tau) }^{\sigma_2(\tau) }d\sigma
\pi_{\mu}(\tau, \sigma) =0~.
\ee
should be valid.
This separation conserves the group of diffeomorphisms  of
the Hamiltonian~\cite{grg} and  leads to the Bergmann-Dirac
generalized action
\be   \label{we3n}
W=\int\limits_{\tau_1}^{\tau_2} d\tau
\left[\left(\int\limits_{\sigma_1(\tau) }^{\sigma_2(\tau) }d\sigma
[ -\pi_{\mu}D_{\tau} \xi^{\mu} -\lambda_1 {\cal H}]\right)
- P_{\mu}\dot X^{\mu} + N_0\frac{P_{\mu}^2}{2{\bar \gamma}} \right] ,~~~
(\bar \gamma=\gamma l(\tau))
\ee
where ${\cal H}$ is the Hamiltonian of local excitations
\be \label{ht}
{\cal H}=
-\frac{1}{2\gamma}[\pi_{\mu}^2+(\gamma \xi'_{\mu})^2]~.
\ee
The variation of the action~(\ref{we3n}) with respect to $\lambda_1$
results in the equation
\be \label{eql1}
\frac{\delta W}{\delta {\lambda_1}}=
 {\cal H}
 - \left(\frac{1}{l {\bar \lambda_1}}\right)^2\frac{P^2}{2 \gamma}=0,
\ee
where
\be \label{46}
{\bar \lambda_1}(\tau,\sigma)=\frac{\lambda_1(\tau,\sigma)}{N_0(\tau)}
\ee
is the reparametrization-invariant component of the local lapse function.
Here we have used the variation of the functional $N_0[\lambda_1]$~(\ref{L})
$$
\frac{\delta N_0[\lambda_1]}{\delta {\lambda_1}}=
\frac{1}{l(\tau){\bar \lambda}_1^2}.
$$
In accordance with our separation of dynamic variables onto the
global and local sectors, the first class constraint~(\ref{eql1})
has two projections onto the global sector (zero Fourier harmonic) and
the local one.
The global part of the constraint~(\ref{eql1}) can be obtained
by variation of the action~(\ref{we3n}) with respect to
$N_0$ (after the substitution of~(\ref{46}) into~(\ref{we3n}))
\be \label{44}
\frac{\delta W}{\delta N_0}= \frac{P^2}{2 \bar \gamma}-H=0,~~~~~~~~
H=\int\limits_{\sigma_1 }^{\sigma_2 } d\sigma
{\bar \lambda_1}{\cal H}~,
\ee
or, in another way, by the integration over $\sigma$
of~(\ref{eql1}) multiplied by $\lambda_1$.
Then, the local part of the constraint~(\ref{eql1})
can be obtained by the substitution of~(\ref{44}) into~(\ref{eql1})
\be \label{50}
{\bar \lambda_1} {\cal H}
 - \frac{1}{l {\bar \lambda_1}}
\int\limits_{\sigma_1 }^{\sigma_2 } d\sigma
{\bar \lambda_1}{\cal H} = 0.
\ee
The integration of the local part over
$\sigma$ is equal to zero
if we take into account the normalization of the local
lapse function
\be \label{norma}
\frac{1}{l(\tau)}\int\limits_{\sigma_1(\tau) }^{\sigma_2(\tau) }
d\sigma \frac{1}{{\bar \lambda_1}} = 1~.
\ee
This follows from the definition of the global lapse function~(\ref{L}).

Finally, we can represent the action~(\ref{we3n}) in the equivalent form
\be   \label{we2}
W=\int\limits_{\tau_1}^{\tau_2} d\tau
\left[\left(\int\limits_{\sigma_1(\tau) }^{\sigma_2(\tau) }d\sigma
[ -\pi_{\mu}D_{\tau} \xi^{\mu}]\right)
- P_{\mu}\dot X^{\mu}- N_0(-\frac{P_{\mu}^2}{2\bar \gamma}
+  H)  \right] ,
\ee
where the global lapse function $N_0$ and
the local one $\bar \lambda_1$ are treated as independent variables, with
taking the normalization~(\ref{norma}) into account after the variation.

According to~(\ref{kt}) and~(\ref{L})
the invariant proper time $T$ measured by the watch of an observer in
the "center of mass" frame of a string is given by the expression
\be \label{wpro}
\sqrt{\gamma}dT:=N_0d\tau,~~~~~~~~~~ \sqrt{\gamma}T=
\int\limits_{0 }^{\tau }d\tau'\left[\frac{1}{l(\tau')}
\int\limits_{\sigma_1(\tau) }^{\sigma_2(\tau) }
d\sigma\frac{1}{\lambda_1(\tau',\sigma)}\right]^{-1}.
\ee
We include the constant $\sqrt{\gamma}$ to provide the dimension of the
time measured by the watch of an observer.

Now we can see from~(\ref{we2}) that the dynamics of the local degrees of
freedom $\pi,\xi$, in the class of functions of nonzero harmonics~(\ref{lhp}),
is described by the same kinemetric invariant
and relativistic covariant equations~(\ref{ex}) where
$x,p$ are changed by $\xi,\pi$, with the set of the first class
(primary and secondary) constraints
\be \label{fcps}
P_{\lambda_1}=0,~~~~~P_{\lambda_2}=0,~~~~~\pi_{\mu}\xi'^{\mu}=0,~~~~~
{\bar \lambda_1} {\cal H}
 - \frac{1}{l {\bar \lambda_1}}
\int\limits_{\sigma_1 }^{\sigma_2 } d\sigma
{\bar \lambda_1}{\cal H} = 0.
\ee
The separation of the "center of mass" (CM) variables
on the level of the action removes the interference terms
which mix the CM variables with the local degrees of freedom;
as a result, the new local constraints~(\ref{fcps})
do not depend on the total momentum $P_{\mu}$,
in contrast to the standard ones.
In other words, there is the problem: when can one
separate the CM coordinates of a relativistic string;
before the variation of the action or after the variation of the action?
The relativistic invariance dictates the first one, because
an observer in the CM frame (which is the preferred frame for a string)
cannot measure the total momentum of the string.

The first class local constraints~(\ref{fcps}) can be supplemented by
the second class constraints
\be \label{scps}
\bar \lambda_1-1=0,~~~~~\lambda_2=0,~~~~~n^{\mu}\xi_{\mu}=0,
~~~~~n^{\mu}\pi_{\mu}=0~,
\ee
where $n_{\mu}$ is an arbitrary time-like vector.
In particular, for $(n_{\mu}=(1,0,0,0))$
the equations of the local constraint-shell action
\be \label{lclc}
W(\mbox{loc.constrs.})=\int\limits_{\tau_1}^{\tau_2} d\tau
\left[\left(\int\limits_{\sigma_1(\tau) }^{\sigma_2(\tau) }d\sigma
\pi_i \dot \xi_i\right)
- P_{\mu}\dot X^{\mu}- N_0(-\frac{P_{\mu}^2}{2\bar \gamma}+ H)\right]
\ee
coincide with the complete set of equations and the same
constraints~(\ref{fcps}),~(\ref{scps}) of the extended action,
i.e., the operations of constraining and variation commute.
The substitution of the global constraint~(\ref{44}) with $\bar \lambda_1=1$
into the action~(\ref{lclc}) leads to the constraint-shell action
\be \label{glclc}
W^D_{\pm}=\int\limits_{X_0(\tau_1)}^{X_0(\tau_2)} dX_0
\left[\left(\int\limits_{\sigma_1(X_0) }^{\sigma_2(X_0) }d\sigma
\pi_i \frac{d\xi_i}{dX_0}\right)
+ P_i\frac{dX_i}{dX_0}\mp \sqrt{P_i^2+2\bar \gamma H}\right].
\ee
This action describes the dynamics of a relativistic string with respect to
the time measured by an observer in the rest frame with the physical
nonzero Hamiltonian of evolution. However, in this system,
equations become nonlinear. To overcome this difficulty,
we pass to the "center of mass" frame.

\section{Levi-Civita geometrical reduction of a string}

To express the dynamics of a relativistic string in terms of
the proper time~(\ref{wpro}) measured by an observer in
the comoving (i.e. "center of mass") frame, we use the Levi-Civita-type
canonical transformations~\cite{lc,gkp1} (as in Section 2.3)
$$
(P_{\mu}, X_{\mu}) \Rightarrow\, (\Pi_{\mu}, Q_{\mu});
$$
they convert the global part of the constraint~(\ref{44})
into a new momentum $\Pi_0$
\be \label{lcct}
\Pi_0= \frac{1}{2{\bar \gamma}}[P_{0}^2 - P_i^2] ,~~~~~~~\Pi_i=P_i,~~~~~~~~~
Q_0=X_0\frac{\bar \gamma}{P_0} ,~~~~~Q_i=X_i-X_0\frac{P_i}{P_0}.
\ee
The inverted form of these transformations is
\be \label{lcif}
P_0=\pm \sqrt{2{\bar \gamma}\Pi_{0}+\Pi_i^2},~~P_i=\Pi_i,~~
X_0=\pm Q_0\frac{\sqrt{2{\bar \gamma}\Pi_{0}+\Pi_i^2}}{{\bar \gamma}} ,
~~X_i=Q_i+Q_0\frac{\Pi_i}{{\bar \gamma}}.
\ee
As a result of transformations~(\ref{lcct}), the
extended action~(\ref{we2})
in terms of the Levi-Civita geometrical
variables takes the form (compare with~(\ref{mec}))
\be   \label{we3}
W=\int\limits_{\tau_1}^{\tau_2} d\tau
\left[\left(\int\limits_{\sigma_1(\tau) }^{\sigma_2(\tau) }d\sigma
[ -\pi_{\mu}D_{\tau} \xi^{\mu}]\right)
- \Pi_{\mu}\dot Q^{\mu}- N_0(-\Pi_0+  H )-\frac{d}{d\tau}(Q_0 \Pi_0)
\right] .
\ee
The Hamiltonian reduction means to resolve
constraint~(\ref{44}) with respect to the momentum $\Pi_0$
\be \label{0cp}
\frac{\delta W}{\delta N_0}=0\,
\Rightarrow\,
\Pi_0= H~.
\ee
The equation of motion for the momentum $\Pi_0$
\be \label{0p}
\frac{\delta W}{\delta \Pi_0}=0\,
\Rightarrow\,
\frac{dQ_0}{d\tau}=N_0~~~~~~~(i.e., dQ_0=N_0d\tau:=\sqrt{\gamma}dT)
\ee
identifies (according to our definition~(\ref{wpro})) the new variable
$Q_0$ with the proper time $T$, whereas the equation for $Q_0$
\be \label{cons}
\frac{\delta W}{\delta Q_0}=0\,
\Rightarrow\,
\frac{d\Pi_0}{d\tau}=0,
~~~~~~~~~~~~i.e.,~\frac{d H}{dT}=0~,
\ee
in view of~(\ref{0cp}), gives us the conservation law.

Thus, resolving the global energy constraint $ \Pi_0=H$, we obtain,
from~(\ref{we3}), the reduced action for a relativistic string in terms of
the proper time $T$
\be   \label{rw3}
W^G=\int\limits_{{T}_1}^{{T}_2} d{T}
\left[\left(\int\limits_{\sigma_1 }^{\sigma_2 }d\sigma
[ -\pi_{\mu}D_{{T}} \xi^{\mu}]\right)
+ \Pi_{i}\frac{dQ_i}{d{T}}- H -
\frac{d}{d{T}} ({T} H)  \right] ,
\ee
where  in analogy with~(\ref{46}) we introduced the factorized "shift-vector"
$\lambda_2=N_0\bar \lambda_2/\sqrt{\gamma}$;
in this case
the covariant derivative~(\ref{cd}) takes the form
\be \label{cd1}
D_{{T}} \xi_{\mu}=\partial_{T}  \xi_{\mu}-
{\bar \lambda_2} \xi'_{\mu}=\frac{D_{\tau}\xi_{\mu}}{N_0}\sqrt{\gamma}~.
\ee
The reduced system~(\ref{rw3}) has trivial solutions for the
global variables $\Pi_i,Q_i$
\be \label{0pi}
\frac{\delta W^R}{\delta \Pi_i}=0\,
\Rightarrow\,
\frac{dQ_i}{d{T}}=0;~Q_i=
\mbox{const};~~~~~~~~~~
\ee
$$
\frac{\delta W^R}{\delta Q_i}=0\,
\Rightarrow\,
\frac{d\Pi_i}{d{T}}=0,~
\Pi_i=\mbox{const}
$$
which have the meaning of initial data.

If the solutions of equations~(\ref{0cp}),~(\ref{0p}), and
~(\ref{0pi}) for the system~(\ref{rw3})
\be \label{geoms}
\Pi_0= H := \frac{M^2}{2 {\bar \gamma}},~~~~
\Pi_i=P_i,~~~~~~~~Q_0={T}\sqrt{\gamma},~~~Q_i=X_i(0),
\ee
are substituted into the inverted Levi-Civita canonical
transformations~(\ref{lcif})
\be \label{rcifs}
P_0=\pm \sqrt{M^2+P_i^2},~~~~~~
X_0({T})= {T}\frac{P_0}{\sqrt{\gamma}l} ~,~~~~~
~~~X_i({T})=Q_i+{T}\frac{P_i}{\sqrt{\gamma}l}~,
\ee
the initial extended action~(\ref{we2}) can be described
in the rest frame of an observer who measures the energy $P_0$ and
the time $X_0$ and sees the rest frame evolution of the "center of mass"
coordinates
\be \label{centre}
X_i(X_0)=Q_i+X_0\frac{P_i}{P_0}~.
\ee
The Lorentz scheme of describing a relativistic
system in terms of the time and energy $(X_0, P_0)$ in the phase space
$P_i,X_i,\pi_{\mu},\xi_{\mu}$
is equivalent to the above-considered the
Levi-Civita scheme in terms of the proper time and  the
evolution Hamiltonian $(T, H)$ in  the phase space
$\Pi_i,Q_i,\pi_{\mu},\xi_{\mu}$,
where the variables $\Pi_i,Q_i$  are cyclic.

\section{Dynamics of the local variables}

\subsection{Reparametrization - invariant reduction for an open string}

We restrict ourselves to an open string with the boundary conditions
\be \label{open}
\sigma_1(T)=0,~~~~~
\sigma_2(T)=\pi,~~~~
~~~~~l(T) =\pi~.
\ee
In the gauge-fixing method, by using the kinemetric transformation,
we can put
\be \label{c12}
\bar \lambda_1=1,~~~~~~~~~~~~~~~~~
\bar \lambda_2=0~.
\ee
This requirement does not contradict the normalization of
$\bar \lambda_1$~(\ref{norma}).

In view of~(\ref{fcps}), it means that
the reduced Hamiltonian  $H$~(\ref{44})
coincides with its density~(\ref{ht})
\be \label{dh}
\bar \phi_{1 }={\cal H}
-\frac{1}{\pi}\int\limits_{0 }^{\pi } d\sigma {\cal H} = 0,~~~~~~~~
\bar \phi_{2 }=\pi_{\mu}\xi'^{\mu}=0
\ee
In this case,
the reparametrization-invariant equations for the local variables
obtained by varying the action~(\ref{rw3})
\be \label{eqx}
\frac{\delta W^R_s}{\delta \xi^{\mu}}=0\,
\Rightarrow\,
\partial_{T}  \pi_{\mu}-
\partial_{\sigma}({\bar \lambda_2}\pi_{\mu})
=\gamma \partial_{\sigma} ({\bar \lambda_1} \xi'_{\mu}),~~~~~
\frac{\delta W^R_s}{\delta \pi^{\mu}}=0\,
\Rightarrow\,
\gamma D_{T}\xi_{\mu}={\bar \lambda_1}\pi_{\mu}
\ee
lead to the D'Alambert equations
\be \label{osci}
   \partial_{T}^2   \xi_{\mu} -
\partial_{\sigma}^2 \xi_{\mu} = 0.
\ee
The general solution of these equations of motion in the class
of functions~(\ref{lhp}) with the boundary conditions~(\ref{open})
is given by the Fourier series
\be \label{smu}
 \xi_{\mu}(T,\sigma)=
\frac{1}{2\sqrt{\pi\gamma}}
[\psi_{\mu}(z_+)+\psi_{\mu}(z_-)],~~~
\psi_{\mu}(z)=i\sum\limits_{n \not= 0 } e^{(-in z)}\frac{\alpha_{n \mu}}{n},~~~
z_{\pm}=T\sqrt{\gamma} \pm \sigma.
\ee
$$
 \xi'_{\mu}(T,\sigma)=
\frac{1}{2\sqrt{\pi\gamma}}
[\psi'_{\mu}(z_+)-\psi'_{\mu}(z_-)],~~~~~~
 \pi_{\mu}(T,\sigma)=
\frac{1}{2}\sqrt{\frac{\gamma}{\pi}}
[\psi'_{\mu}(z_+)+\psi'_{\mu}(z_-)]~.
$$
The total coordinates $Q_{\mu}^{(0)}$ and momenta $P_{\mu}$ are
determined by the reduced dynamics of the "center of mass"~(\ref{0pi}),
~(\ref{geoms}),~(\ref{rcifs}), and the string mass $M$ obtained from~(\ref{44})
\be \label{p0m}
P_{\mu}^2=M^2=2\pi\gamma H=
2\pi\gamma\int\limits_{0 }^{\pi }d\sigma {\cal H}.
\ee
The substitution of $\xi_{\mu}$ and $\pi_{\mu}$ from~(\ref{smu})
into~(\ref{ht}) leads to the Hamiltonian density
$$
{\cal H}=-\frac{1}{4\pi}
\left[\psi'^2_{\mu}(z_+)+
\psi'^2_{\mu}(z_-)\right]~,
$$
and from~(\ref{p0m}) we obtain, for the mass, the expression
\be \label{mass}
M^2=-2\pi\gamma \bar L_0~=-\frac{\gamma}{2} \int\limits_{0 }^{\pi }
d\sigma \left[(\psi'_{\mu}(z_+))^2+(\psi'_{\mu}(z_-))^2\right]~.
\ee

The second constraint~(\ref{dh}) in terms of the vector
$\psi'_{\mu}$ in~(\ref{smu}) takes the form
\be \label{gsc}
\xi'_{\mu}\pi^{\mu}=\frac{1}{4\pi}
\left[\psi'^2_{\mu}(z_+)-
\psi'^2_{\mu}(z_-)\right]=0~\Rightarrow~~
\psi'^2_{\mu}(z_+)=
\psi'^2_{\mu}(z_-)=\mbox{const.}~,
\ee
and the first constraint~(\ref{dh}) $\bar \phi_1=0$ is satisfied identically.
After the substitution of the constant value~(\ref{gsc})
into~(\ref{mass}) we obtain
that $\mbox{const.}=-M^2/\pi\gamma$; thus, finally
the reparamerization-invariant constraint takes the form
\be \label{cm}
P^2_{\mu}+ \pi\gamma \psi'^2_{\mu}(z_{\pm})=0 ~~~~~~ (~P^2_{\mu}= M^2~)~.
\ee
Unlike this constraint, the gauge-fixing
reparametrization-noninvariant constraint~\cite{gsw,bn}
\be \label{cc}
\left(P_{\mu}+ \sqrt{\pi\gamma} \psi'_{\mu}\right)^2=0
\ee
contains the interference of the local and global degrees of freedom
$\psi'_{\mu}P^{\mu}$.
The latter violates
the relativistic invariance of the local excitations which form
the mass and spin of a string.

The constraint~(\ref{cm}) in terms of the Fourier components~(\ref{smu})
takes the form
\be \label{fcm}
\psi'^2_{\mu}(z_{\pm})=
\sum\limits_{k,m \not= 0}^{ }\alpha_{k,\mu}\alpha_{m}^{\mu}
e^{-i(k+m)z_{\pm}}=
2\sum\limits_{n=-\infty }^{\infty }
\bar L_ne^{-inz_{\pm}}=-\frac{M^2}{\pi\gamma}~,
\ee
where $\bar L_n$ are the contributions of the nonzero harmonics
\be \label{nzf}
\bar L_0 =
-\frac{1}{2}\sum\limits_{k \not= 0}\alpha_{k \mu} \alpha^{\mu}_{-k}~,
~~~
\bar L_{n \not= 0}=-\frac{1}{2}\sum\limits_{k \not= 0,n }
\alpha_{k \mu } \alpha^{\mu}_{n-k}~.
\ee
From~(\ref{fcm})
one can see that the zero harmonic of this constraint determines the mass
of a string
\be \label{0zh}
M^2=-2\pi\gamma \bar L_{0}= -\pi\gamma
\sum\limits_{k \not= 0 } \alpha_{k \mu } \alpha_{-k \mu}
\ee
and coincides with the gauge-fixing value. However, the nonzero harmonics
of constraint~(\ref{fcm})
\be \label{vabl}
\bar L_{n \not= 0}=
-\frac{1}{2}\sum\limits_{k \not= 0,n } \alpha_{k \mu } \alpha_{n-k \mu}=0,~~~~
\bar L_{-n} =\bar L_n^*
\ee
(as we dicussed above) strongly differ from the nonzero harmonics of the
gauge-fixing constraints~(\ref{cc}).
The latter (in the contrast to~(\ref{cm})) contains the mixing the
global motion of the center of mass $P_{\mu}$ with the local
excitations $\psi_{\mu}$. It is clear that this mixing the
global and local motions violates the the Poincare invariance of
the local degrees of freedom.

The algebra of the local constraints~(\ref{vabl}) of
the reparametrization-invariant dynamics of a relativistic string
is not closed, as it does
not contains the zero Fourier harmonic of the energy constraint
(which has been resolved  to express the dynamic equations in terms of
the proper time).

The reparametrization-invariant dynamics of a relativistic
string in the form of the first and second class constraints
~(\ref{fcps}),~(\ref{scps}) coincides with the R\"ohrlich approach to
the string theory~\cite{roh}.
This approach is based on the choice of
the gauge condition
$$
p_{\mu}\xi^{\mu}=0,
p_{\mu}\pi^{\mu}=0~~~~ \Rightarrow~~~~~~
G_n=P_{\mu}\alpha_n^{\mu}=0,~~~~~~n \not= 0~,
$$
instead of (\ref{scps}). As consequence of this gauge the constraints
(\ref{cm}), (\ref{cc}) became equivalent. In quantum theory,
this condition is used for eliminating the states with negative
norm in the "center of mass" (CM) frame
(in our scheme, the CM frame appears as a result of
the geometric Levi-Civita reduction).
This reference frame is the only preferred frame for quantizing such a
composite relativistic object as the string, as only in this
frame one can quantize the initial data. This is a strong version of the
principle of correspondence with classical theory: the classical
initial data become the quantum numbers of quantum theory.

\subsection{Quantum theory}

Thus, our classical Hamiltonian reparametrization - invariant formalism
provides the quantization of the string as in R\"ohlich gauge.

The R\"ohrlich approach distinguishes two cases: $M^2=0$ and $M^2\not= 0$.

The first case, in our scheme, the equality $M^2=0$ together
with the local constraints~(\ref{vabl}) form the Virasoro algebra.
The reparametrization-invariant version of the Virasoro algebra
(with all its difficulties, including
the $D=26$ - problem and the negative norm states) appears only in
the case of the massless string $-2\pi\gamma\bar L_0=M^2=0$.

In the second case $M^2\not= 0$, the R\"ohrlich gauge $\alpha_{n,0}=0$.
allows us to exclude the time Fourier components
$\alpha_{n0}$, and it is just these components that after quantization
$$
[\alpha_{n,\mu},\alpha_{n,\nu}^+]=-m \eta_{\mu,\nu} \delta_{m,n};~~~~
(n,m\not= 0,~~\eta_{00}=-\eta_{ii}=1)
$$
lead
to the states with negative norm because of the system being unstable.
This means that the state vectors in the CM frame are constructed only by
the action on vacuum of the spatial components of the operators
$a^+_{ni}=\alpha_{-n i}/\sqrt{n},n>0$ \cite{roh}
\be \label{fi}
|{\bf \Phi}_{{\bf \nu}}>_{CM}
=\prod\limits_{n=1 }^{\infty }
\frac{(a^+_{nx})^{\nu_{nx}}}{\sqrt{\nu_{nx}!}}
\frac{(a^+_{ny})^{\nu_{ny}}}{\sqrt{\nu_{ny}!}}
\frac{(a^+_{nz})^{\nu_{nz}}}{\sqrt{\nu_{nz}!}}|0>~,
\ee
where the three-dimensional vectors ${\bf \nu}_n=(\nu_{nx},\nu_{ny},\nu_{nz})$
have only nonnegative integers as components.
These state vectors automatically satisfy the constraint
\be \label{a0}
\alpha_{n0}|{\bf \Phi}_{{\bf \nu}}>_{CM}=0,~~~~~n > 0
\ee
The physical states~(\ref{fi}) are subjected to further constraints
(\ref{vabl}) with $n \geq 0$
\be \label{vabl0}
\bar L_{n}|{\bf \Phi}_{{\bf \nu}}>_{CM}=0,~~~~~n > 0,
~~~~P^2=M^2_{{\bf \nu}}=\pi\gamma<{\bf \Phi}_{{\bf \nu}}\sum\limits_{m\not= 0 }^{ }
\alpha_{-m,i}\alpha_{m,i}|{\bf \Phi}_{{\bf \nu}}>~,
\ee
where $\bar L_n$ can be represented in the normal ordering form
\be \label{m2f}
\bar L_{n>0}=\sum\limits_{k= 1 }^{\infty }\alpha^+_{k,i}\alpha_{n+k,i} +
\frac{1}{2} \sum\limits_{k>1 }^{n-1 }\alpha_{k,i}\alpha_{n-k,i}.
\ee
Constraints (\ref{a0}) and (\ref{vabl0}) are the
first class constraints, in accordance with the Dirac
classification~\cite{d2}
because they form a closed algebra for $n,m >0$
\be \label{lgbr}
 [G_n,G_m]=0,~~~[\bar L_n,\bar L_m]=(n-m)\bar L_{n+m},~~~
[G_n,\bar L_m]=nG_{m+n}~.
\ee
Therefore the conditions~(\ref{a0}) eliminating the ghosts and the
conditions~(\ref{vabl0}) defining the physical vector states are consistent.
Note that the commutator $[\bar L_n,\bar L_m]$  does not contain a c-number
since suffices $n \geq 0$ and $m \geq 0$ in the Virasoso operators
$\bar L_n$ do not lead to the central term.

On the operator level, equations determining
the resolution of the constraints are fulfilled in a weak sense, as only the
" annihilation" part of the constraints is imposed on the state vectors.

In quantum theory, one can introduce a complete set of eigen functions
satisfing equations
\be \label{88}
H[\pi_i,\xi_i] <\xi|{\bf \nu}>=\frac{M^2_{{\bf \nu}}}{2\pi\gamma}
<\xi|{\bf \nu}>,
\ee
where
$$
<\xi|{\bf \nu}>=<\xi|{\bf \Phi_{\nu}}>,~~~~~~~~
\sum\limits_{\bf \nu }^{ }<\xi_1|{\bf \nu}><{\bf \nu}|\xi_2>
=\prod\limits_{\sigma }^{ }
\delta^3(\xi_1-\xi_2).
$$

\section{The causal Green functions}

Now we can construct the causal Green function for a relativistic string
as the analogy of the causal Green function for a relativistic
particle~(\ref{Dl}) -~(\ref{srfi}) discussed in Section 2.

The Veneziano-type causal Green function
is the spectral series with the Hermite polynomials
$<\xi|\nu>$ over the physical state vectors
$|{\bf \Phi}_{\bf \nu}>=|{\bf \nu}>$
\be \label{ven}
G_c(X|\xi_1,\xi_2)=G_+(X|\xi_1,\xi_2)\theta(X_0)+
G_-(X|\xi_1,\xi_2)\theta(-X_0)=
\ee
$$
i\int\limits_{ }^{ }\frac{d^4P}{(2\pi)^4}\exp(-iPX)\sum\limits_{\nu }^{ }
\frac{<\xi_1|\nu><\nu|\xi_2>}{P^2-M^2_{\nu}-i\epsilon}.
$$
The commutative Green function for a relativistic string $G_+(X|\xi_1,\xi_2)$
can be represented in the form of the
Faddeev-Popov functional integral~\cite{fp}
in the local gauge~(\ref{scps})
\be \label{strfi}
G_+(X|\xi_2,\xi_1)=\int\limits_{X(\tau_1)=0 }^{X(\tau_2)=X }
\frac{dN_0(\tau_2)d^4P(\tau_2)}{(2\pi)^3}
\prod\limits_{\tau_1 \leq \tau < \tau_2}^{ }\left\{ d\bar N_0(\tau)
\prod\limits_{\mu}\left( \frac{dP_{\mu}(\tau)dX_{\mu}(\tau)}{2\pi}\right)
\right\}
F_+(\xi_2,\xi_1),
\ee
using the representation of the spectral series
\be \label{ssps}
F_+(\xi_2,\xi_1)=\sum\limits_{{\bf \nu} }^{ }
<\xi_2|{{\bf \nu}}>\exp\left\{ i W[P,X,N_0,M_{\bf \nu}]\right\}
<{\bf \nu}|\xi_1>=
\ee
in the form of the functional integral
$$
F_+(\xi_2,\xi_1)=
\int\limits_{\xi_1 }^{\xi_2 }
D(\xi,\pi) \Delta_{fp}
\exp\left\{ i W_{fp}\right\}~,
$$
$W[P,X,N_0,M_{\bf \nu}]$ is the action~(\ref{SR}) with the
mass $M_{\nu}$
\be \label{fp1}
W_{fp}=\int\limits_{0}^{\tau(X_0)} d\tau \left[ -
\left(\int\limits_{0 }^{\pi }d\sigma
\pi_{\mu}\dot \xi^{\mu}\right)- P_{\mu}\dot X^{\mu}-{N_0}
\left(-\frac{P^2}{2\pi\gamma}+ H\right)\right]
\ee
is the constraint-shell action~(\ref{lclc}),
\be \label{prod}
D(\xi,\pi)=
\prod\limits_{\tau,\sigma}
\prod\limits_{\mu }^{ }\frac{d\xi_{\mu} d\pi_{\mu}}{2\pi}~,
\ee
and
\be \label{fps}
\Delta_{fp}=
\prod\limits_{\tau,\sigma}
\delta(\phi_1)) \delta(\pi_0)
\delta(\phi_2)) \delta(\xi_0)
det B^{-1},~~~~~det B=det\{\phi_1,\phi_2,\pi_0,\xi_0\}
\ee
is  the FP determinant
given in the monograph~\cite{hrt}.

\section{Conclusion}

To describe the invariant dynamics of constrained
relativistic string  we
used the universal method of the Hamiltonian reduction of
their actions by resolving the energy
constraint,
so that one of variables of the extended phase space
(with a negative contribution to the energy constraint) converts
into the invariant evolution parameter,
and its conjugate momentum becomes the invariant Hamiltonian of
evolution.

This method allows us to find integrals of motion by the Levi-Civita
canonical transformations
which converts the energy constraint into a new momentum, and the time-like
variable of the world space into the proper time interval.
For a particle and a string the Levi-Civita transformations are the
Hamiltonian form of the Lorentz transformations
which describe pure relativistic effects of the transition  from
the rest frame of reference to the comoving one.

We have shown that a relativistic string can be described directly
in terms of the
reparametrization-invariant parameter of evolution with the nonzero
Hamiltonians of evolution
in agreement with the equations of motion of the initial system.

A crucial point in our approach is
the separation of the "center of mass" coordinates on the level of the action.
The definition of the proper time with the nonzero Hamiltonian of
evolution consistent with the
group of diffeomorphisms of the Hamiltonian description requires to
separate the "center of mass" coordinates before varying the action,
whereas in the standard gauge-fixing method, the "center of mass" coordinates
are separated after varying the action.
The operations of separation of the "center of mass" coordinates
and variation of the action do not commute.
The relativistic invariance dictates the reparametrization - invariant
way,  as an observer in the comoving frame cannot measure the
componets of the total momentum of a string.
Unique admissible gauge is the R\"ohrlich gauge that leads directly
to the quantum theory of a string without a critical dimension.

Thus, we can formulate the novelty of this work:
i) the separation of the "center of mass" coordinates on the level
of the action, ii) finding of the integrals of motion by the Levi-Civita
transformation, iii) deriving of the nonzero Hamiltonian of evolution
of a string with respect to the proper time with the new algebra
of the Poisson brackets, that provides the R\"orlich gauge, and
iv) constructing  of new  reparametrization - invariant path
integral representations of the causal Green functions for
relativistic particle and  string.

{\bf Acknowledgments}

\medskip

We are happy to acknowledge interesting and critical
discussions with H. Kleinert, E. A. Kuraev,
V. V. Nesterenko, A. I. Pashnev, and I. V. Tyitin.

\vspace{1cm}

{\Large\bf Appendix A: Kinemetric transformations}

\vspace{0.5cm}


The kinemetric transformations of the differentials
$$
\widetilde{\tau} = \dot f_1(\tau)d\tau,~~~~~~~~~d\widetilde{\sigma}=
\dot f_2(\tau,\sigma)d\tau + f_2'(\tau,\sigma)d\sigma
$$
correspond to transformations of the string coordinates
$$
x_{\mu}(\tau,\sigma)=\widetilde{x}_{\mu}(\widetilde{\tau}),
\widetilde{\sigma}),~~~~~~~~~~~~
x'_{\mu}(\tau,\sigma)=
\widetilde{x}'_{\mu}(\widetilde{\tau},\widetilde{\sigma})f'_2(\tau,\sigma),
$$
$$
\dot x_{\mu}(\tau,\sigma)=
\dot {\widetilde{x}}_{\mu}(\widetilde{\tau},\widetilde{\sigma})\dot f_1(\tau)+
\widetilde{x}'_{\mu}(\widetilde{\tau},\widetilde{\sigma})\dot f_2(\tau,\sigma),
$$
>From these equations, we can derive the transformation law
for $\lambda_1, \lambda_2$ taking into account~(\ref{lambda2})
$$
\lambda_1(\tau,\sigma)=
\frac{\sqrt{(\dot x x')^2 - \dot x^2 x'^2}}{x'^2(\tau,\sigma)}=
\frac{\sqrt{({\dot {\widetilde{x}}} \widetilde{x}')^2 -
{\dot {\widetilde{x}}}^2 \widetilde{x}'^2}}{\widetilde{x}'^2(\widetilde{\tau},\widetilde{\sigma})}
\frac{\dot f_1}{f'_2}=\widetilde{\lambda}_1 \frac{\dot f_1(\tau)}{f'_2(\tau,\sigma)}.
$$
$$
\lambda_2(\tau,\sigma)=\frac{\dot x x'}{x'^2}=
\frac{(\dot {\widetilde{x}} \widetilde{x}')\dot f_1f'_2+ \widetilde{x}'^2\dot f_2f'_2}{\widetilde{x}'^2f'^2_2}=
\widetilde{\lambda}_2\frac{\dot f_1}{f'_2}+\frac{\dot f_2}{f'_2}.
$$
The kinemetric-invariance of the interval~(\ref{m}) with respect to~(\ref{kt})
follows from these transformation laws
and the transformation of the conformal factor
$$
\Omega(\tau,\sigma)=f'_2(\tau,\sigma)\widetilde{\Omega}(\widetilde{\tau},\widetilde{\sigma})
$$
The covariant derivative~(\ref{cd}) is transformed under~(\ref{kt}) as
$$
D_{\tau}x_{\mu}=
\dot x_{\mu}- \lambda_2 x'_{\mu}=
\dot f_1(\tau)\left[\dot {\widetilde{x}}_{\mu}-
{\widetilde{\lambda}}_2 {\widetilde{x}}'_{\mu}\right]=
\dot f_1(\tau)D_{\widetilde{\tau}} \widetilde{x}_{\mu}~.
$$

\end{document}